\newcount\subeqnno
\def\beginsubeqn{\begingroup
\refstepcounter{equation}
\subeqnno=\arabic{equation}
\setcounter{equation}{0}
\def\theequation{\the\subeqnno\alph{equation}}
}
\def\endsubeqn{\setcounter{equation}{\the\subeqnno}\endgroup}

\newcommand{\uu}[1]{\verb!#1!\endgroup}

\newcommand{\mb}[1]{\ifmmode#1\else\mbox{$#1$}\fi}

\newcommand\al{\mb{\alpha}}

\newcommand\de{\mb{\delta}}
\newcommand\ep{\mb{\epsilon}}

\newcommand\et{\mb{\eta}}
\newcommand\th{\mb{\theta}}

\newcommand\la{\mb{\lambda}}

\newcommand\rh{\mb{\rho}}
\newcommand\si{\mb{\sigma}}

\newcommand\ph{\mb{\phi}}

\newcommand\Ga{\mb{\Gamma}}
\newcommand\De{\mb{\Delta}}

\newcommand\Ph{\mb{\Phi}}

\newcommand\calL{\mb{{\cal L}}}

\newcommand{\x}{\mb{\times}}

\newcommand{\beq}{\begin{equation}}
\newcommand{\eeq}{\end{equation}}
\newcommand{\nn}{\nonumber}

\newcommand{\bea}{\begin{eqnarray}}
\newcommand{\eea}{\end{eqnarray}}
 
\newcommand{\fn}{\footnote}
\newcommand{\norm}[1]{\parallel \! {#1} \! \parallel}
\newcommand{\mod}[1]{\mid \! {#1} \! \mid}

\newcommand{\emb}{\mb{{\rm emb}}}
\newcommand{\Ad}{\mb{\rm Ad}}
\newcommand{\ad}{\mb{\rm ad}}
\newcommand{\tr}{\mb{\rm tr}}
\newcommand{\deriv}[2]{\frac{d {#1}}{d {#2}}}

\documentstyle[12pt]{article}
\topmargin = - 0.5 cm
\begin{document}
\bibliographystyle{unsrt}
\include{epsf}


\begin{flushright}
DAMTP-95-48
\end{flushright}
\begin{center}
\LARGE{Topological Inflation, without the Topology\\}
\vspace*{0.5cm}
\large{Nathan\  F.\  Lepora
\fn{e-mail: N.F.Lepora@damtp.cam.ac.uk}
and Adrian Martin
\fn{e-mail: A.Martin@damtp.cam.ac.uk}}\\
\vspace*{0.2cm}
{\small\em Department of Applied Mathematics and Theoretical
Physics,\\
University of Cambridge, Silver Street,\\
Cambridge, CB3 9EW, U.\ K.\\ }
\vspace*{0.2cm}

{January 1996}
\end{center}%

\begin{abstract}
We extend the `topological inflation' of Linde and Vilenkin to {\em
unstable} monopoles. This allows the monopole to decay; not
inflating eternally, as topological inflation demands. Such a
situation happens naturally in some Grand Unified Theories --- such as
supersymmetric flipped-$SU(5)$. We analyse analytically the dynamics
of inflating monopoles to determine the equations governing the
expansion, additionally recovering the bound on the scale of symmetry
breaking found numerically by Sakai {\em et. al.} The latter half of
this paper is devoted to the Cosmology of inflating unstable monopoles
--- which is an example of an inhomogeneous cosmology. We describe how
such a monopole may be formed and how long it inflates for --- finding
it to be a random process. We then derive how cosmological parameters,
such as density and temperature, are distributed at the end of
inflation, and how the Universe reheats as the monopole decays. The
general conclusion of this work is that such inflation creates a local
region of relatively flat, homogenous and isotropic Universe
surrounded by pre-GUT matter.

\end{abstract}
\thispagestyle{empty}
\newpage
\setcounter{page}{1}


\section{Introduction.}
\label{sec 1}

Inflation is commonly accepted to be an essential ingredient of
standard big bang cosmology. This acceptance is motivated by an
overwhelming amount of (circumstantial) evidence. It is
hard to conceive of another mechanism solving so many of the
unexplained features of non-inflationary big bang cosmology: Why is
our observable universe so flat? Why is it so isotropic and
homogeneous? How did the galaxies clump? Why does all we see seem to
have been in causal contact? And so on. In addition,
inflation seems to fit naturally in quantum cosmological versions of
creation; these predict that the universe should collapse 
quickly (Planck times), and inflation gives a mechanism for making the
Universe large, as we see today. 

The standard way to introduce inflation is through vacuum domination:
a large energy density of vacuum produces gravitational effects 
causing exponential expansion of the distance scale. Because the vacuum 
does not scale with expansion (unlike energy density and temperature,
which are diluted) this process continues until some other 
effect takes over. 

However, stopping the above process is difficult --- this is referred
to as the `graceful exit problem'. In addition, since the original
matter has been diluted away, one needs creation of all matter
seen today. A sensible postulate is that the inflationary
exit must somehow convert vacuum energy into particles and temperature
(which don't drive inflation) --- this is called `reheating'.

A sensible scenario for implementing the above was the `old
inflationary universe' scenario of A.~Guth \cite{Guth81}. This linked
in nicely with the theory of phase transitions in Grand Unified
Theories (GUT's): in passing through the GUT phase transition the
Universe remains in a long-lived metastable state (by virtue of GUT's
being first order) of false vacuum, this drives the inflation and
then termination proceeds via quantum tunnelling into true
vacuum. This scenario is very attractive because of its naturalness: it
seems fairly likely that GUT's exist and this scenario can be a
consequence of them. Unfortunately, this scenario does not work:
large inhomogeneities created by tunnelling out of the
inflationary state are incompatible with observations of homogeneity,
such as the cosmic microwave background.

After the old inflationary scenario, modern theories of inflation 
make use of additions to the minimal field theory; 
contrasting with the old inflation scenario where one {\em
  uses what is given}. Also, an important part of modern inflationary
scenarios are the concepts of `chaotic inflation', where the correct
initial conditions (provided the field theory is right) are guaranteed
from the ensemble of possible Universes.

However, recently Linde \cite{Lind94} and Vilenkin \cite{Lind94} have
put forward a 
novel new way to implement inflationary expansion of the Universe:
topological inflation. In this scenario they show that, providing
conditions are correct, the core of a topological defect may undergo
exponential inflationary expansion. This scenario was further
elucidated in a paper by A. D. Linde and D. A. Linde \cite{Lind94b}.
Again, as for the `old
inflationary scenario', such a scenario is natural because one is
using GUT phase transitions to invoke inflation: one is using 
something that is given anyway. 

The central idea of topological inflation is very simple:
restored symmetry in 
the core of a topological defect gives energy to the vacuum, which
drives inflation of the core of the defect. One expects that the
vacuum energy would have to be sufficiently high, and Vilenkin and
Linde claim that the condition is that the scale of symmetry breaking
\[
\eta > O(m_{\rm pl}).
\]
The inflation that ensues is eternal, since the defect may never
decay. In addition, quantum fluctuations create a fractal nature of
encrusted defects.

They have several well motivated reasons for expecting gravitational
effects to become important in the centre of a defect when the
scale of symmetry breaking is near Planck scale:
\begin{itemize}
\item
The size of the false vacuum region (the core of the defect) is
greater than the order of the Horizon size corresponding to the vacuum
energy at the centre of the defect (the inverse of the calculated
Hubble parameter)
\item
The Schwarzschild radius, corresponding to the mass of the defect, is
greater then the order of the core radius of the defect.
\item
For global defects the deficit angle due to gravitational lensing
becomes greater than $2\pi$.
\item
The usual slow-roll condition on inflation that the potential $V(r)$
must be such that
\[
\left| \frac{V'}{V} \right| < \frac{\sqrt{48\pi}}{m_{\rm pl}},
\]
implies that for a potential of Landau type $\la(\phi^2 -\et^2)^2$ the
scale of symmetry breaking must be as above. Naively this condition
seems the most tentative, but actually it proves to be the most
important.
\end{itemize}

In addition simulations of global defects coupled to gravity, by Sakai
{\em et. al.} \cite{Saka95}, confirm the above; additionally yielding
the more exact relation:
\[
\et > \frac{m_{\rm pl}}{4}.
\]

With topological inflation the monopole inflates eternally,
since the core is stabilised topolgically and it is the restored
symmetry in the core that provides the vacuum energy for inflation.
One may terminate inflation {\em locally} by living on the edge of the
core; inflation continues eternally in the centre of the monopole,
whilst at the edge the gauge fields are inflated away, destabilising
the scalar field --- this scalar field thus decays terminating the
inflation. This has the effect of creating space and matter in the
centre of the core and pushing it outwards. Such a situation would
produce an inhomogenous cosmology on large scales across the region
that has terminated inflation: nearer the core (the place where
matter is created) less time has passed since that region terminated
inflation, whilst further away form the core more time has
passed. Therefore as one travels towards the core (across the region
where inflation has terminated) the Universe becomes hotter, more
dense, {\em etc.} Such large scale inhomogeneities could be small,
depending upon the dynamics of the decay. However, small scale
inhomogeneities, {\em i.e.} density perturbations, created from a
necessarily close to Planck scale process are large.

Additionally, within the inflating region quantum creation of
further monopoles may happen, which also inflate. One thus ends with 
a fractal nature of monopoles, with monopole inside monopoles inside
.... All of which inflate.

We wish to contrast with this situation by using unstable field
configurations, for example embedded monopoles.  In 
terms of the fields that make up the configuration {\em there is no
  difference between a topological monopole and an unstable embedded
  monopole}, hence unstable embedded monopoles may inflate too. The
amount of inflation achieved will depend upon how long the monopole
takes to decay, but we shall show later that this duration is {\em
  probabilistic}. Hence, one only needs {\em one} such monopole to
randomly live long enough to create what we see today.

Thus in the scenario sketched here, the core of the monopole decays,
not inflating eternally. Furthermore, it seems unlikely that a fractal
nature will ensue from quantum creation of embedded monopoles within
the inflating region: we will show later that the probability of a
monopole `living' long enough to inflate is extremely small.

Thus in the scenario sketched here, an embedded monopole is created
randomly in a long lived inflationary state, it inflates and then
decays. The result being a creation of a region of homogenous,
isotropic, flat space bounded by GUT scale physics. Because there is
more inflation at the centre of the core than on the edge (the vacuum
energy is larger) the boundary between GUT physics and the inflated
region is `squeezed' to be very sharp.

The plan of this paper is as follows. Section (\ref{sec 2}) is a
brief review of embedded monopoles; covering definition, properties,
and existence. This serves to set the scene for section ~(\ref{sec
  3}), 
where we couple monopole configurations to gravity facilitating 
investigation of their inflationary properties: we derive when
inflation may take place, recovering the expression of Sakai, {\em
  et. al.}, and determine the equations controlling inflationary
expansion. These inflationary equations are used in section (\ref{sec
  4}) 
to derive the cosmology --- which is inhomogeneous. We extensively
connect monopole inflation to the facets of the standard inflationary
scenario. Thus section (\ref{sec 4}) is necessarily broad in scope, and
we take a chronological view for presentation: beginning with 
quantum cosmology, then passing through a discussion of unstable
monopole creation 
and time for decay, onto their inflationary expansion and decay, which
reheats the universe (discussed in both fast and slow reheat
scenarios), finally ending with the termination of our universe back
into the GUT epoch. We conclude our discussions with section (\ref{sec
  5}). 


\section{Embedded Monopoles}
\label{sec 2}

This section is devoted to a brief review of the existence and
properties of 
embedded monopoles. This sets the scene for coupling such
solutions to gravity for an investigation of their inflationary
properties. 

We first define an embedded monopole solution, with relation to its
unstable nature. Then we show how such solutions may be found in GUT's
--- {\em i.e.} which GUT's admit embedded monopoles as solutions, and
whether  
other defects are also admitted. To illustrate we consider
the GUT flipped-$SU(5)$ --- we expect this to be the best situation
for the inflationary scenario presented;
correspondingly, we give details on the parameters of that theory,
which we shall use later in sec. (\ref{sec 4}) when discussing
cosmology. Finally, we review the Prasad-Sommerfield approximation
to the monopole`s profile.

\subsection{What is an Embedded Monopole?}

The shortest answer to this question is: it is a monopole solution
that is not topologically stable --- in the sense of Vachaspati, {\em
  et. al.} \cite{Vach94}.

To answer the question in more depth, one needs to quickly review the
archetypal topologically stable monopole solution \cite{t'Hooft76}
--- namely, that in
$SU(2) \rightarrow U(1)$ under an adjoint representation of Higgs. The
Lagrangian density for that theory is:
\beginsubeqn
\label{su2 lagrangian}
\bea
\calL &=& \frac{1}{2} \left\{ (D_\mu \Ph)^\dagger (D^\mu \Ph) \right\} -
\frac{1}{4} \tr(F_{\mu \nu} F^{\mu \nu}) - \la \left( \tr (\Ph^\dagger \Ph)
- \et^2 \right)^2 \\
{\rm with}\ \ 
D_\mu \Ph &=& \partial_\mu \Ph - q \, \ad(A_\mu) \Ph.
\eea
\endsubeqn
Here the Higgs field $\Ph$ and gauge field $A^\mu$ take values in
$L(SU(2))$ --- for which we use the usual Pauli spin basis $\{
\frac{i}{2} \si^a \}$. Also, $\ad(A_\mu) \Ph$ refers to the adjoint
action of $L(SU(2))$ on $L(SU(2))$; namely, $\ad(A_\mu) \Ph = [ A_\mu,
\Ph]$.

The magnitude of $\Ph$ is fixed by minimisation of the potential,
breaking the symmetry. However, the direction of $\Ph$ is not fixed
--- with the degeneracy given by the orientations of the broken $U(1)$
in $SU(2)$ --- which is $SU(2)/U(1) \cong S^2$. In three spatial
dimensions, boundary conditions for configurations are defined upon
the two-sphere at infinity. Thus, we may regard definition of the
boundary conditions in this case as a map $S^2 \rightarrow S^2$ ---
which admits non-trivial boundary conditions that may not be deformed
continuously to triviality. This solution is the magnetic monopole:
\beginsubeqn
\bea
\underline{\Ph}(\underline{r}) &=& f_{\rm mon}(r) \underline{\hat{r}},
\\ 
A^\mu (\underline{r}) &=& \frac{g_{\rm mon}(r)}{r} \ep_{\mu a b}
\frac{i \si^a}{2},
\eea
\endsubeqn
where we are treating $\underline{\Ph}$ to be a vector within
$L(SU(2))$. The above configuration is imbued stability through the
topological nature of the boundary conditions; since the boundary
conditions may not be deformed to triviality, the monopole may not
decay (classically).

An embedded monopole is derived from this basic solution. Consider a
general symmetry breaking $G \rightarrow H$, with condensation of a
Higgs field $\Ph$. Providing the group theory allows (see later for
clarification), one may choose a subtheory ( the embedded subtheory)
which acts like the $SU(2) \rightarrow U(1)$ theory above. By
extending a monopole solution on the subtheory back to the full theory
(by making other fields vanish) one obtains the embedded monopole
\cite{Vach94}.

However, in extending the monopole solution back to the full theory,
one of two situations may happen as regarding stability:
\begin{itemize}
\item The embedded monopole loses its topological nature --- this is
  what we shall refer to as an `embedded monopole'. These are
  unstable.
\item The monopole keeps its topological nature --- we shall refer to
  these as topological monopoles.
\end{itemize}
To distinguish, one must examine the topology of $G/H$. Provided $G/H$
is connected then one has topological monopoles if the second
homotopy group $\pi_2(G/H)$ is non-trivial.

\subsection{How do we find Embedded Monopoles?}

Since we shall be using unstable embedded monopoles to cause
inflation, it is important to know which GUT's admit them. Not all
GUT's do --- many only having topologically stable monopoles. Hence, we
shall describe a formalism \cite{me2} for finding embedded monopole
solutions.

This prescription for finding embedded defects rests upon
the observation that one may define a specific defect from generators 
in the Lie algebra. Explicitly using this correspondence allows one to
determine the family structure and stability properties of the set of
embedded solutions. The prescription is group theoretic in nature;
one writes the Lie algebra of $G$ as:
\beq
LG = LH \oplus \underline{m},
\eeq
interpreting the vector space $\underline{m}$ as the set of generators
generating massive gauge bosons. Correspondingly, it is
$\underline{m}$ which generates embedded defect solutions to the
model. Group theoretic considerations identify the sets
of generators that generate gauge inequivalent families of embedded
defects to be the irreducible spaces of $\underline{m}$ under the
adjoint action of $H$. Namely, write
\beq
\underline{m}=\underline{m}_1 \oplus \cdots \oplus \underline{m}_N,
\eeq
with $\underline{m}_i$ irreducible under the transformation of $T \in
\underline{m}_i$ given by $T \mapsto {\Ad}(h)T = hTh^{-1}$, for $h \in
H$. Then embedded defect solutions are generated by elements in
$\underline{m}_i$.

Embedded monopole solutions are of the form:
\beginsubeqn
\label{monopole}
\bea
\Ph(r,\th) &=& f_{\rm mon}(r) \underline{\hat{r}},\\
A(r,\th) &=& \frac{g_{\rm mon}(r)}{r} \ep_{\mu a b} T_b,
\eea
\endsubeqn
with $\underline{\hat{r}}$ a radial vector in the corresponding Higgs'
embedded subtheory. The generators of the monopole solution are $T_1,
T_2 \in \underline{m}_2$, with the constraints: $T_3 = [T_1,T_2] \in
LH$; the inner product $\langle T_1, T_2 \rangle = 0$; $\norm{T_1} =
\norm{T_2}$; and $D(e^{2 \pi T_1}) \Ph_c = \Ph_c$.

Provided $G/H$ is connected, monopole solutions are unstable if the
second homotopy group of $G/H$ is trivial.

A couple of examples of theories that admit embedded monopole
solutions have been discussed in the literature. The notable example
is flipped-$SU(5)$, which we shall discuss next. Another example of a
theory which has unstable embedded monopole solutions is Pati-Salam
$SU^4(4)$, see \cite{me3} for a discussion of these.

\subsection{Example: Flipped-$SU(5)$}

For a more detailed discussion of embedded defects and their
properties in flipped-$SU(5)$, see \cite{me1}.

The gauge group is $G=SU(5) \x \widetilde{U(1)}$ \cite{Barr82}, which
acts on a ten 
dimensional, complex Higgs field (conveniently represented
as a five by five, complex antisymmetric matrix) by the {\bf
  10}-antisymmetric representation. Denoting the generators of $SU(5)$
as $T^a$ and $\widetilde{U(1)}$ as $\widetilde{T}$, the derived
representation acts on the Higgs field as:
\beq
d(A_i^a T^a + \widetilde{A}_i \widetilde{ T}) 
= g \al^i (T^i \Ph + \Ph {T^i}^\top) + \tilde{g} \al^0 \widetilde{T}
\Ph.
\eeq
Here $g$ and $\tilde{g}$ are the $SU(5)$ and $\widetilde{U(1)}$
coupling constants, respectively. The corresponding representation of
the Lie algebra is the exponentiation of this.

It is necessary for the 
following discussion to know a couple of generators explicitly,
namely:
\beq
T^{15} = i \sqrt{\frac{3}{10}} 
\left( \begin{array}{ccc}  \frac{2}{3} \bf{1}_3 & \vdots & \bf{0} \\ 
\cdots & \cdots & \cdots \\
\bf{0} & \vdots & - \bf{1}_2
\end{array} \right),\ \ \ \ 
\widetilde{T} = i \sqrt{\frac{12}{5}} \bf{1}_5.
\eeq
These generators are properly normalised with respect to a standard
inner product on the Lie algebra.

For a vacuum given by
\beq
\Phi_c = {v \over \sqrt{2}} 
\left( \begin{array}{ccc} \bf{0}_3 & \vdots & \bf{0} \\ 
\cdots & \cdots & \cdots \\
\bf{0} & \vdots & I
\end{array} \right),\ \ \ {\rm where}\ \ \ 
I = 
\left( \begin{array}{cc} 0 & 1 \\ 
-1 & 0
\end{array} \right),
\eeq
one obtains breaking to the standard model $H = SU(3)_c \x SU(2)_I \x
U(1)_Y$ provided the parameters of the potential satisfy $\et^2, \la_1
>0$ and $(2 \la_1 + \la_2) > 0$.
The V and hypercharge fields, and their generators, are given by 
\beginsubeqn
\bea
V_i = \cos \Theta A_i^{15} - \sin \Theta \widetilde{A}_i, & &
T_V = \cos \Theta T^{15} - \sin \Theta \widetilde{T},\\
Y_i = \sin \Theta A_i^{15} + \cos \Theta \widetilde{A}_i, & &
T_Y = \sin \Theta T^{15} + \cos \Theta \widetilde{T}.
\eea
\endsubeqn
where the GUT mixing angle is $\tan \Theta = \widetilde{g}/g$. Then $d(LH)
\Ph_{\rm vac} = 0$, with the isospin and colour symmetry groups
nestled in $SU(5)$ as
\beq
\left( \begin{array}{ccc}  SU(3)_c & \vdots & \bf{0} \\ 
\cdots & \cdots & \cdots \\
\bf{0} & \vdots & SU(2)_I
\end{array} \right) \subset SU(5).
\eeq

Then, to find the embedded defect spectrum one determines the
reduction of $LG$ into $LG = LH \oplus \underline{m}$ and finds the
irreducible 
spaces of $\underline{m}$ under the adjoint action of $H$. The space
$\underline{m}$ reduces 
into two irreducible spaces under the adjoint action of $H$, which
are:
\beq
\label{m}
\underline{m}_1 = \{\al T_V: \al \ {\rm real}\},\ \ \ \ \ \ 
\underline{m}_2 = 
\left( \begin{array}{ccc}  \bf{0}_3 & \vdots & \underline{A} \\ 
\cdots & \cdots & \cdots \\
-\underline{A}^\dagger & \vdots & \bf{0}_2
\end{array} \right).
\eeq

It is the generators in the space $\underline{m}_2$ which define
embedded 
monopole solutions of the form eq. (\ref{monopole}). 
An example of such a pair is: $T_1$ defined from
$\{\underline{v}_1^\top = (i,0), \underline{v}_2 = \underline{v}_3 =
0\}$, and $T_2$ defined from $\{\underline{v}_1^{\prime \top} = (0,i),
\underline{v}_2^{\prime \top} = \underline{v}_3^{\prime \top} = 0\}$.

Flipped-$SU(5)$ admits no topologically stable monopole solutions;
only 
embedded monopoles. The reason for this is that the second homotopy
group of the vacuum manifold is trivial. Namely
\beq
\pi_2 \left( \frac{G}{H} \right) = \frac{\pi_1(H)}{\pi_1(G)} =
\frac{\pi_1(U(1)_Y)}{\pi_1(\widetilde{U(1)})} = \bf{1}.
\eeq
The last step is not as trivial as it first looks ---
it depends upon the projection of $\widetilde{U(1)}$ onto $U(1)_Y$
being non-trivial.

Our situation for inflation, which we shall describe later, should
work for other non-topological defects as well. Thus, as regarding 
monopole type configurations, it is interesting to speculate whether a
Sphaleron \cite{Mant83} type configuration could be used --- we show in the
appendix that flipped-$SU(5)$ admits no Sphaleron solutions.

It is necessary to discuss the possible parameter values in 
flipped-$SU(5)$. The stringent requirement being equality of 
isospin and colour gauge couplings at unification. The running of the
coupling constants is generally dependant 
upon whether we are considering the supersymmetric version of the
theory or not. 

In the non-supersymmetric case the hypercharge gauge coupling does not
equal the values of the colour and isospin gauge couplings at
unification. Estimates from the running of gauge couplings (see
\cite{Lang93}) yield at unification the following quantities: 
an energy scale of $10^{16}$ to $10^{17}$ GeV; isospin
and colour gauge couplings of about $1/50$; and a hypercharge gauge
coupling of about $1/37$.

For the supersymmetric case one has considerable freedom in the choice
of the breaking scale of super-symmetry and hence in the unification
scale and values of the gauge couplings. However, in superstring
motivated flipped-$SU(5)$ the coupling constants are required to meet
\cite{Lope93} --- this yields a breaking scale of about $10^{18}$ GeV.

\subsection{The Prasad-Sommerfield Approximation}
\label{sec 2.4}

For later sections of this paper, namely cosmological applications, it
is necessary to know the profile functions $f_{\rm mon}(r)$ and
$g_{\rm mon}(r)$ of the monopole. These may be conveniently
approximated for part of the parameter space by the Prasad-Sommerfield
approximation \cite{Pras75}.

In the, so called, Prasad-Summerfield limit, where $\sqrt{\la}/g
\rightarrow 0$, then the profile functions tend towards:
\beginsubeqn
\label{ps limit}
\bea
f_{\rm mon}(r) &\rightarrow& f_{\rm PS}(r) = \et \left( {\rm
  coth}(\frac{r}{r_0}) - \frac{r_0}{r} \right) \\
g_{\rm mon}(r) &\rightarrow& g_{\rm PS}(r) = \left( \frac{r}{r_0}
\right) {\rm sinh}(\frac{r}{r_0})^{-1},
\eea
\endsubeqn
where $r_0 = (g \et)^{-1}$.

Examination of the validity of the Prassad-Sommerfield approximations
is discussed in \cite{Kirk81}. They conclude: in
the near region (i.e the core) the size of the core and the profiles
are not significantly altered for non-zero $\sqrt{\la}/g$; it is the
asymptotic behaviour that is effected.

Since we shall be interested in the core of the monopole, we shall use
the Prassad-Summerfield approximation whenever necessary.


\section{Monopole Inflation}
\label{sec 3}

This section is intended to be generally applicable to both
topological and embedded monopoles. After summarising the field
theory and monopole solution that we deal with, we derive when a
(topological or embedded) monopole solution may inflate. 
We shall also highlight the root
of the differences by showing how the topology of the theory
gives rise to eternal inflation for topological monopoles, but not for
embedded monopoles.

We wish to determine when a monopole configuration will undergo
exponential inflation. To determine this one 
minimally couples (as in eq.~(\ref{action}), below) the monopole
configuration of sec. (\ref{sec 2}) to gravity. 

Our philosophy in deriving the conditions on whether inflation takes
place is to seek an (approximate) steady state 
solution of these equations. Then from validity of these equation one
determines where in the parameter space inflation occurs.

Since the embedded monopole is not
topologically stable it decays. Provided that it takes a
sufficiently long time to decay (see later on this point) enough
e-folds of inflation occur to be compatible with present day
observations of flatness, {\em etc.}

\subsection{The Field Theory}

The field theory we shall be dealing with is stipulated to be of
Yang-Mills type and admits monopole solutions (embedded and/or
topological). In general we shall require it to be physically
motivated --- {\em i.e.} a well motivated Grand Unified Theory.

The Lagrangian density describing the interaction of a Higgs field
$\Phi$ coupled to a gauge field $A_\mu$ is of the form in eq.
(\ref{su2 lagrangian}), but generalised to a general group $G$
\beginsubeqn
\bea
\calL &=& \frac{1}{2} \left\{ (D_\mu \Ph)^\dagger (D^\mu \Ph) \right\} -
\frac{1}{4} \tr(F_{\mu \nu} F^{\mu \nu}) - V(\Ph), \\
{\rm with}\ \ 
D_\mu \Ph &=& \partial_\mu \Ph - q \, d(A_\mu) \Ph.
\eea
\endsubeqn
Here $q$ represents the charge corresponding to the derived
representation $d$ of the Higgs field (which is a map from $LG$ to
actions on $\Phi$ and describes how the gauge field acts on $\Phi$).
In general the Higgs potential for a Grand Unified Theory is
necessarily complicated by $\tr(\Phi^4)$ terms, though we shall assume
(for simplicity) that we are dealing with a purely Landau type
potential, 
\beq
V(\Phi) = \la \left( \tr (\Ph^\dagger \Ph) - \et^2 \right)^2.
\eeq
Conclusions are not expected to change for more complicated
potentials.

This theory is stipulated to have a monopole solution (topological or
embedded) and, using the techniques of the previous section, write
it as eq. (\ref{monopole}):
\beginsubeqn
\bea
\Ph(r,\th) &=& f_{\rm mon}(r) \underline{\hat{r}},\\
A(r,\th) &=& \frac{g_{\rm mon}(r)}{r} \ep_{\mu a b} T_b,
\eea
\endsubeqn

\subsection{The Inflationary Solution of Yang-Mills-Einstein Theory}

The action representing the minimal coupling of gravity to a field
theory, defined by a Lagrangian $\calL$, is
\beq
\label{action}
S_{\Phi} = \int d^{4}x\sqrt{-g}\cal L
\eeq
where $g$ is the determinant of the metric tensor.

We shall assume {\em isotropy} of the metric tensor. This assumption
is equivalent to assuming that the monopole is created in a
spherically symmetric state and decays through the zeroth angular
momentum decay mode. Hence, we shall take the metric tensor to be
\beq
g_{ab}=\mbox{diag}
\left(-1,a^{2}(r,t),a^{2}(r,t),a^{2}(r,t)\right),
\eeq
where $a(r,t)$ is the usual scale factor --- which relates the
physical distance scale $r'$ (at time $t$) to the initial distance
scale $r$ by 
\beq
r'(t) = \int^r_0 a(x,t) {\rm d}x.
\eeq
It should be noted that the expected derived cosmology
shall be {\em isotropic} about the
centre of the monopole, but {\em not} homogenous in space; thus the
scale factor depends upon distance as well as time.
This should be contrasted with Friedman-Robertson-Walker
Cosmologies which assume {\em both} homogeneity as well as isotropy.
In addition the Hubble parameter will also depend upon distance from
the origin, as well as time, though it is defined in the usual fashion:
\beq
H(r,t) = \frac{\dot{a}(r,t)}{a(r,t)}.
\eeq

The equations of motion of our Yang-Mills-Einstein theory are
obtained by variation with respect to $\Phi$ and $\sqrt{-g}$.
The direction of $\Phi$ in Higgs-space becomes irrelevant, so we shall 
denote $\phi = \norm{\Phi}$ (precisely because we are assuming that the
initial configuration and decay are spherically
symmetric). Stationarity of variation yields the field equations: 
\beginsubeqn
\label{Phieq} 
\bea
\ddot{\phi}+3H\dot{\phi}+\Gamma\dot{\phi}+\Omega\phi &=&
-\frac{dV}{d\phi}\\
H^{2}\equiv\frac{8\pi}{3m_{\rm pl}^{2}}T_{00}&=&
\frac{8\pi}{3m_{\rm pl}^{2}}\left(V(\phi)+
\frac{1}{2}\dot{\phi}^{2}+a^{-2}(\underline{\nabla}\phi)^{2}\right).
\eea
\endsubeqn
where
\begin{eqnarray}
\Gamma=-2A_{0} &,& 
\Omega=[D_{i}D^{i}-(\partial_{0}A^{0})+A_{0}A^{0}-3HA_{0}].
\end{eqnarray}
We now make some comments on the above equations. 


We should comment on interpretation. In the first equation 
the expansion of the Universe manifests itself as a damping term
proportional to the Hubble parameter, whilst 
the term $\Lambda(A) \dot{\phi}$ represents the interaction between gauge
fields and the rate of decay of the embedded monopole --- one can
think of the configuration outside the core as trying to `pull' the
core off the peak of the Higgs potential. 

However, the important observation is that, apart from some spatial
gradient and gauge terms, the above equations are {\em the same as}
the usual equations governing slow-roll inflation with a scalar
potential $V(\phi)$. Coupled with the fact that inflation {\em
  dilutes} spatial gradients and gauge fields to zero, it is now
becoming clear how to proceed: we are looking for an
(approximately) steady state solution that is inflating, so we
examine the field equation when gradients and gauge fields are diluted
away. 

Hence, write the Higgs field for the decaying (through the zeroth
angular momentum mode)monopole configuration as
\beginsubeqn
\label{fields}
\bea
\Ph (r,\th,\varphi,t) = \ph(r,t) \underline{\hat{r}} \\
{\rm with}\ \ \ \ \ \ph(r,0) = \et f_{\rm mon}(r).
\eea
\endsubeqn
Remember that for a monopole configuration $A_{0}=0$, so $\Gamma=0$
and $\Omega\Phi=-a^{2}\nabla^{2}\Phi$.

Ignoring spatial gradients and gauge fields, using (\ref{fields}) and
treating $\phi$ as a classical field, the field equations become
\beginsubeqn
\label{field equations}
\bea
\ddot{\phi}+3H\dot{\phi}+V^{\prime}(\phi) &=& 0
\label{phieq}\\
H^{2} &=& \frac{8\pi}{3m_{\rm pl}^{2}}\left(V(\phi)+
\frac{1}{2}\dot{\phi}^{2}\right).
\eea
\endsubeqn
These equations are, of course, the usual slow-roll equations governing
inflation. 

Inflation occurs in the usual way: providing $\ddot{\ph} \ll 3H
\ph$ then inflation happens whenever the potential is sufficiently
flat to satisfy (these equation may be easily derived from
eq.~(\ref{field equations}) ---
see \cite{Kolb}, for instance)
\beginsubeqn
\bea
\mod{V^{\prime \prime}} &\leq& 9H^2,\\
\left|\frac{V' m_{\rm pl}}{V}\right| &\leq& \sqrt{48 \pi}.
\eea
\label{inflcond}
\endsubeqn
In homogenous inflationary scenarios these equations  serve to act as
a restriction only upon the parameters of the theory; in the
inhomogeneous scenario considered here they also act as a restriction
upon the region of space that may undergo inflation.
One interprets the potential energy of the false vacuum at
the centre of the monopole as driving the core to
inflate, where the potential is sufficient to do so. 

We now determine the parameter space that allows inflation from
the above two conditions.

Initially we take $\dot{\phi}=0$
\footnote{In practice, since monopole
configurations are formed in collisions involving four or more bubbles 
during the phase transition, $\dot{\phi}$ will not always be zero
initially. 
However, for inflation to occur we still need $\dot{\phi}^{2}<
V(\phi)$ and so we take $\dot{\phi}=0$ as an estimate here. Also if
$\dot{\phi}\neq0$ initially we will have non-zero $A_{0}$ which
will tighten the first inflation constraint via eq. (\ref{Phieq}).} 
so that
\beq
H^{2} =\frac{8\pi}{3m_{\rm pl}^{2}}V(\phi)
\eeq
and our first condition for inflation, eq.~(\ref{inflcond}a), becomes
\beq
|V^{\prime\prime}|  \leq \left(\frac{24\pi}{m_{\rm pl}^2}\right)V.
\eeq
Setting $\phi^{2}=\et^{2}f^{2}$, this can be re-expressed as
\beq
\left| f^2_{\rm mon}-\frac{1}{3} \right| \leq
2\frac{\pi}{\lambda} \left( \frac{\et}{m_{\rm pl}} \right)^2
(f_{\rm mon}^2-1)^2
\label{criteq}
\eeq
where $f_{\rm mon}$ is the monopole profile function.

\begin{figure}[t]
\vspace{3in}
\includegraphics{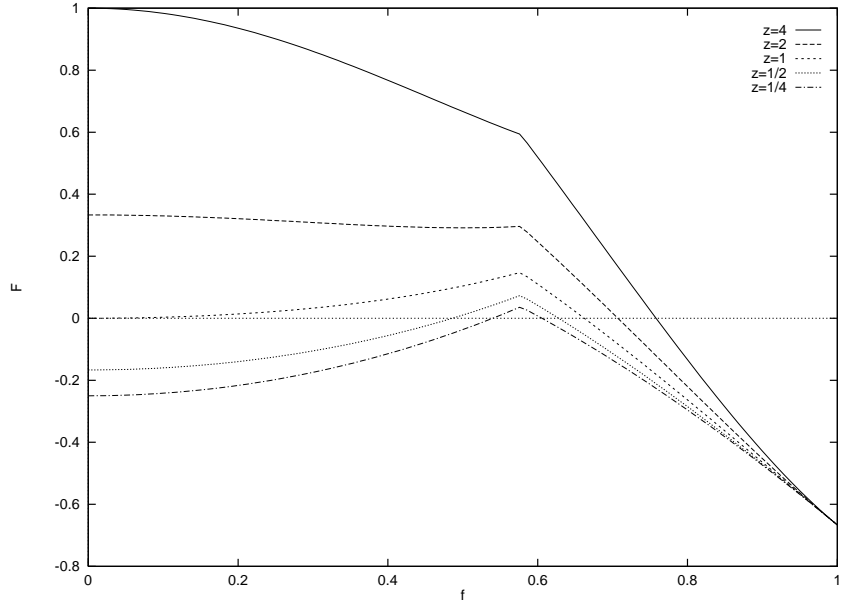}
\caption{$F(f)=
      2\pi\Lambda(\frac{\eta}{m_{\rm pl}})^2(f^2-1)^2-
      |f^2-\frac{1}{3}|$ versus $f$ for three values of 
      $z=6\pi\Lambda(\frac{\eta}{m_{\rm pl}})^2$. Inflation is
      possible in the regions where $F>0$. The edge of the
      defect core is at $f=1$.}
\label{fig1}
\end{figure}

From eq.~(\ref{criteq}) and fig. (1) we see that if
\begin{equation}
\left( \frac{\et}{m_{\rm pl}} \right)^2\geq
\frac{1}{6\pi}
\label{constr1}
\end{equation} 
then inflation is triggered at the centre of the core. 

Examining fig. (1) you may be concerned that for
$|\phi|=\et/\sqrt{3}$ inflation always occurs.
We have neglected the second constraint however. 
It is straightforward to show using the second that inflation is only
possible at $|\phi|=\eta/\sqrt{3}$ if 
\beq
  \left(\frac{\et}{m_{\rm pl}}\right)^2  > \frac{1}{4\pi}
\label{constraint}
\eeq
and so if it occurs there, by the first condition it occurs for all
$|\phi|<\eta/\sqrt{3}$.

In conclusion of this section: we conclude that inflation takes place
in the core of the monopole provided that eq.~(\ref{constraint}) is
satisfied. 

Before we discuss how the above condition may be realised physically,
we shall quickly mention another feature that restricts theoretical
parameters. If we approximate the energy density, $\rho$, inside 
the core as $V(\phi)$ then it is straightforward to show
that the ratio of the Schwarzschild radius to the core radius is
\beq
  \frac{r_{\rm schwarzschild}}{r_{core}}=
  \frac{1}{2}\lambda\left(\frac{\eta}{m_{\rm pl}}\right)^2.
\label{constr2}
\eeq
This condition constrains $\la$, from whether or not the 
inflating region shall be cut off from the rest of the Universe by an
event horizon. When the Schwarzschild radius is larger than the
core radius, then the monopole will be shielded to an external
observer who may see the monopole   
as a Reissner-Nordstrom black-hole \cite{Lind94}. In addition
this may revive the eternal inflation feature of topological inflation ---
by stabilising the monopole-like boundary condition at the horizon, so
that the embedded monopole can not decay.

\subsection{Satisfying the Parameter Space}

For physical Grand Unified Theories the constraint
\[
 \left(\frac{\et}{m_{\rm pl}}\right)^2  > \frac{1}{4\pi}
\] 
seems to require a rather high value for the scale of symmetry
breaking. We now describe two physical situations in which such scales
should be realised.

\subsubsection{Fundamental String Inspired Theories}

In many superstring derived theories it is expected that unification
happens at the superstring unification scale, which is of order $10^{18}$
to $10^{19}$GeV \cite{Anto91}. These energies are sufficient to cause
inflation in the core of the defect. This is favourable
for our scenario as it is generally believed that supersymmetry is
pretty much essential for discussing GUT scale physics because of the
gauge hierarchy problem, and superstrings seem to be a sensible
scenario for this. Therefore we require a sensible string motivated
theory that admits unstable monopole solutions.

The most favourable situation for this seems to be fundamental string
derived flipped-$SU(5)$ \cite{Anto87}. In this it is necessary to
introduce several extra extra mass scales which raise the usual
supersymmetric unification point ($10^{16}$ to $10^{17}$GeV) to
superstring unification energies (about $10^{18}$ to
$10^{19}$GeV) \cite{Lope94}. These extra mass scales seem to be well
motivated, and 
their actual origin would lie within the structure of the string
model.

\subsubsection{Interpretation of the Running Couplings}

We make this suggestion rather tentatively. The conclusions of this
suggestion are, however, rather successful.

In general the scale of symmetry breaking for a Grand Unified Theory
is found from analysis of the running coupling constants --- the
region where they come together yields an estimation of the energy of
Unification. Technically, what one is plotting in these graphs are the
coupling constants against the centre of mass energy of collisions
between two particles.

To make numerical estimates of the scale of symmetry breaking one
takes this centre of mass energy to be of the same order as a typical
gauge boson, $m_X^2 \sim g \et^2$ to yield a value for $\et$. This
should give a good {\em order of magnitude} estimation for the scale,
but for more accurate determinations it is difficult to see why the
centre of mass energy at the phase transition and the typical mass of a
gauge boson should be the same.

Instead we postulate that the
centre of mass energy (at the phase transition) is better estimated by
the critical temperature of the phase transition. The critical
temperature of a phase transition is of the form \cite{Lind74}:
\beq
\label{T_c}
T_{\rm c}^{2} \sim  \frac{\et^2}{a + b q^{2}/2\lambda},
\eeq
with the constants $a,b \sim O(1)$ depending upon the details of the
theory, and $q$ is the value of the coupling constants at unification
(we are assuming that the coupling constants unify, though conclusions
are qualitatively unchanged if they do not) which is typically of
order $10^{-2}$.

Hence if we take $T_{\rm c} \sim 10^{16}-10^{17} {\rm GeV}$, then we
may satisfy the condition in eq.~(\ref{T_c}) by fine tuning $\la$ ---
requiring $\la$ to be less than about $10^{-5}$. Coincidentally, we
shall find this value is also required by many of the cosmological
constraints on the model.

\subsection{Decay of the Monopole and Links to Topology}

The equation eq.~(\ref{field equations}) describes how the scalar field
and 
gravitational field evolve. This equation gives information as to how
the gravitational metric evolves --- describing the distance scaling
and hence inflation. Also contained within this equation is
information as to how the monopole decays, by studying the evolution
of the scalar field. 

Eq.~(\ref{field equations}) describes the evolution of the scalar
field through 
inflation; describing also the decay of the monopole. Neglecting terms
due to gauge fields, which are diluted away, 
eq.~(\ref{field equations}) is 
\beq
\ddot{\ph} + 3 H \dot{\ph} = -\deriv{V}{\ph},
\eeq
strictly speaking $\Ph$ is the magnitude of the Higgs field. Assuming
that the embedded monopole has been created in a long lived stationary
state is equivalent to requiring
\beq
\ddot{\ph} \ll 3 H \dot{\ph}.
\eeq
Making this approximation yields the equation governing the
(slow-roll) decay of $\ph$; which is
\beq
\label{phi dot}
\dot{\ph} = \frac{\la \et^2}{H_0} \ph.
\eeq
From this equation we may glean some information
about how topological nature relates to eternal inflation.

Recall that the arguments in this section have been general to both
embedded and topological monopoles; arguments so far relying only on
the form of the fields. The important point about a
topological monopole is that because the boundary conditions are not
deformable to triviality then somewhere there {\em must} be a zero of the
Higgs field; this argument being invalid for embedded
monopoles. In fact, for embedded monopoles it is to be expected that
nowhere is the Higgs field {\em exactly} zero. This is how the
topology related to the dynamics: if at some point $p$ we have
$\phi(p)=0$ then by eq. (\ref{phi dot}) this point may never
decay. Hence, 
since for a topological monopole somewhere the Higgs field has to
vanish then a topological monopole has the eternal inflation feature;
embedded monopoles don't.


\section{Cosmological Implications}
\label{sec 4}

In this section we derive some of the cosmological effects of the
mechanism derived in the previous chapter. Because of the intrinsic
inhomogeneity of the inflating monopoles, the cosmology derived from
such is an {\em inhomogeneous} cosmology. As an
explicit example we derive in detail how an inhomogeneous cosmology
may fit into `the standard lore'.

We order this section chronologically. Showing how the model of
inflating unstable monopoles may realistically give rise to a Universe
such as we see today. Starting by outlining how this model may fit
into principles of quantum cosmology, we then go onto discuss how such
a monopole may be formed and the conditions created afterwards. 
Making such conditions compatible with what we see today serves to
constrain the model, and we explicitly derive some of these
constraints. We finish this section by giving a necessary consequence
of such an inhomogeneous inflationary scenario, namely a different
version of the ultimate fate of the Universe.

It should be noted that the version of inflation presented here
necessarily solves the `graceful exit problem'.

\subsection{The Beginning : Quantum Cosmology}

It is generally believed that the Universe spontaneously nucleates out
of nothing, giving an ensemble of possible, randomly initialised
Universes. We live in an element of this ensemble that is able to
evolve to conditions such as we see today. Chaotic inflation assumes a
necessary ingredient of this scenario to be creation in a state of
vacuum domination --- which fuels the inflation that is necessary to
go from a tiny region to our large Universe. Although the
ensemble is dominated by Universes not of this form; the probability
distribution for our initial conditions is conditional (on us
existing), and therefore that must have been the initial state.

Our scenario may be incorporated into this picture quite
naturally. Instead of stipulating that a necessary initial condition
of the Universe must be vacuum domination large enough to give all the
necessary inflation, we instead postulate creation in a GUT,
containing embedded monopoles, with sufficient expansion to reach the
GUT phase transition. Then an embedded monopole created at this phase
transition may achieve the rest of the necessary inflation.

As we shall explain later, the amount of inflation achieved by an
embedded monopole is probabilistic --- determined by the
initial state of the monopole after formation. Although a formation of
an embedded monopole in such a state is unlikely, only one is required
to achieve conditions today. To weight the expectation, so that such a
situation may happen, a combination of the following two criteria is
required:
\begin{itemize}
\item Some amount of pre-expansion, so that many embedded monopoles are
  formed.
\item Many nucleations of Universes similar to the required initial
  state. 
\end{itemize}

Principles of Quantum Cosmology attempt to prescribe a probabilistic
weight to the initial conditions of the Universe. To properly compare
whether embedded monopole inflation is more likely than chaotic
inflation one must properly take account of the above two
conditions. We do not attempt such a calculation. However, we may
compare supersymmetric (SUSY) GUT's to non-supersymmetric (NON-SUSY)
GUT's via a simple argument. Since the energy that a Universe will
cool to, before re-collapse, is strongly biased towards higher (i.e
Planck scale) energies, one would expect SUSY GUT's to be more
probable because of their higher unification scales (other features
not withstanding). This bias towards SUSY GUT's follows for many other
features of the cosmology, as we shall explain later.

\subsection{The Onset of Inflation: at Unification}
\label{sec 4.2}

Provided the Universe is created with sufficient expansion to reach
the GUT phase transition, the symmetry breaks and defects are formed
via the Kibble mechanism \cite{Kibb76}. All GUT phase transitions are 
strongly first order \cite{Lind74}, thus the transition completes by
bubble nucleation. We are assuming that the GUT contains embedded
monopoles as 
part of its defect spectrum; of which flipped-$SU(5)$ is the notable
example, though there are others --- such as Pati-Salam $SU^4(4)$.

We sketch to the reader how embedded monopoles are formed in such a
situation. Inside each bubble the Higgs field takes a phase. When four
bubbles collide and coalesce, interpolation over the four relative
phases in space creates a Higgs-gauge configuration, which, provided
the core is in the symmetric phase, is gauge inequivalent to the
vacuum. Then the time evolution of such a configuration depends upon
the exact details of the initial configuration 
(which is determined by the timing, orientation and phase of
the colliding bubbles).

There is a nice mental picture of the classical time evolution of such
a configuration. Picture the configuration space (which consists of
the field components and conjugate momenta to the fields) graphed
against the configuration's energy --- this forms a surface
within the energy-configuration space. An embedded monopole
configuration (eq.~(\ref{monopole})) is the centre of a saddle point
within 
that surface. Now, when the four bubbles collide and a configuration is
formed, one can think of it as a random point on this surface with a
certain random amount of velocity, and then the classical evolution of
that configuration is determined by
rolling along the surface. The ultimate destination
being a minimum on the surface, which, providing the 
theory has no topological defects, corresponds to the trivial vacuum.

This picture is useful in determining the amount of inflation from an 
embedded monopole. As we shall show in the next section, the number of
e-folds of inflation is roughly proportional to the lifetime of the
configuration, and cosmologically, one needs several phase
transition times of life. However, the expected time before an
embedded monopole decays is of order one phase transition
time. This problem is resolved by realising that the decay time is
in fact probabilistic, and, although the probability of a bubble
collision giving successful inflation is small, providing enough
collisions take place it should happen --- after all, we only need
{\em one} such event to produce a Universe like ours. Such a
configuration is created exactly on the saddle point, or, perhaps
moving towards the saddle with just the right amount of velocity. One
can liken the probability of such a configuration being created from a
four bubble collision as akin to shooting three snooker balls at each
other and the collision stopping them dead.

Guessing that the probability distribution of the decay time of the
defect is exponentially tailed, 
with a mean about one phase transition time (for small times), the
distribution should be something like
\beq
\label{probability}
P(t_{decay}>t) = e^{-\frac{t}{\al}},
\eeq
where $\al$ is the {\em expected} decay time, for small times 
a sensible estimate is $\al \sim {\et}^{-1}$. We shall argue later
at larger times the parameter $\al$ increases.

\subsection{Steady State Inflation}
\label{sec 4.3}

After the embedded monopole has started inflating it rapidly reaches
an era of steady state inflation, where the spatial curvature inside
the monopole may be neglected. The subsequent evolution within the
monopole can be modelled well quite simply until the embedded
monopole decays and inflation ceases.

We shall use $r$ as a radial coordinate to describe distances, where
$r$ was a physical distance before the onset of inflation. From
eq. (\ref{monopole}) the Higgs field takes the values $\phi(r) =
f_{\rm mon}(r) \ph_c$, with $f_{\rm mon}(r)$ being the monopole
profile function --- which is conveniently approximated by the
Prasad-Sommerfield limit for parameter ranges that we shall
consider (see sec.~(\ref{sec 2.4})). By substituting this into the
Landau 
potential we may determine the vacuum energy at distance $r$ to be:
\beq
\label{vacuum energy}
V(r) = \la \left( \ph^\dagger \ph - \et^2 \right)^2 
     = \la \et^4 \left( f_{\rm mon}(r)^2 -1 \right)^2.
\eeq
This decreases monotonically from $\la \et^4$ at $r=0$, to zero at
infinity. This variation in the vacuum energy will produce a variable
Hubble parameter with distance, and hence a variation in the number of
e-folds.

To obtain the variation of scale factor $a(r)$ with distance one could
calculate the Einstein equations with a vacuum energy $V(r)$
above. This is clearly a very complicated calculation. Instead we
shall make a couple of approximations to simplify the
analysis. Firstly, we shall neglect any time evolution of $\ph(r)$ ---
so these calculations will not deal with decay. Secondly, we shall
neglect spatial derivatives in the scale factor --- this is expected to
be a good approximation inside the core and at large distances where
the fields have little gradient; also inflation rapidly dilutes
spatial gradients due to expansion of space.

Making these two approximations is equivalent to dealing with a local
Friedmann-Robertson-Walker (FRW) cosmology, described locally by the
FRW equations, but over large 
distances the parameters change. Hence the equations describing the
evolution of the scale factor are the usual FRW equations, but with a
variable vacuum energy of the form in eq.~(\ref{vacuum energy}) ---
this yields: 
\beginsubeqn
\bea 
\left(\frac{\dot{a}}{a}\right)^2 + \frac{k}{a^2} &=& \frac{8 \pi}{3
    m_{\rm pl}^2} V(r), \\
2 \frac{\ddot{a}}{a} + \left(\frac{\dot{a}}{a}\right)^2 +
\frac{k}{a^2} &=& \frac{8 \pi}{ m_{\rm pl}^2}V(r).
\eea
\endsubeqn
These equations are self consistent and yield a solution only if $k =
0$; which shows they are only a valid approximation once the monopole
has passed its initial transient phase of inflation. The solution is
easily obtained to be
\beq
\label{scale factor}
a(r,t) = a_0 \exp \left( H_0 t (1- f_{\rm mon}(r)^2) \right),
\ \ {\rm with}\ \ H_0 = \sqrt{\frac{8 \pi \la \et^4}{3 m_{\rm pl}^2}}.
\eeq 
Here $a_0$ is the initial scale factor before inflation, which is
arbritrary; serving only as a reference. One interprets the scale
factor $a(r,t)$ as 
meaning: after a time $t$ of inflation the physical radial
coordinate $r'$ is related to $r$ by the usual scaling relation
\beq
\label{scaling}
r'(r,t) = \int_{x=0}^{r} \frac{a(x,t)}{a_0} {\rm d}x.
\eeq
Also, the Hubble parameter at time $t$ of inflation, at a distance
$r$, is easily calculated to be
\beq
\label{hubble parameter}
H(r,t) = \frac{\dot{a}}{a} = H_0 \left(1 - f_{\rm mon}(r)^2 \right),
\eeq
which is constant in $t$ --- as expected. When one is talking
about parameters which vary with distance one should use the physical
coordinate $r'$ to describe them; hence, we shall plot quantities
parametrically against $r'$. Figs. (2) and (3) illustrate variation of
the Hubble parameter with distance and time --- fig. (2) being a
snapshot of the Hubble parameter at times $1/H_0, 2/H_0, 3/H_0$, and
fig. (3) a snapshot at times $10/H_0, 11/H_0, 12/H_0$. These figures
illustrate well the exponential expansion of the core and
homogenisation of the Hubble parameter within the core. The cores
exponential expansion is expected from inflation. The homogenisation
of the Hubble parameter has also a simple explanation: because the
vacuum energy is higher in the centre of the core than at the edge,
more inflation happens at the centre, which effectively
pushes parameter variations to the edge of the inflated core. 

\begin{figure}[p]
\vspace{3in}
\includegraphics{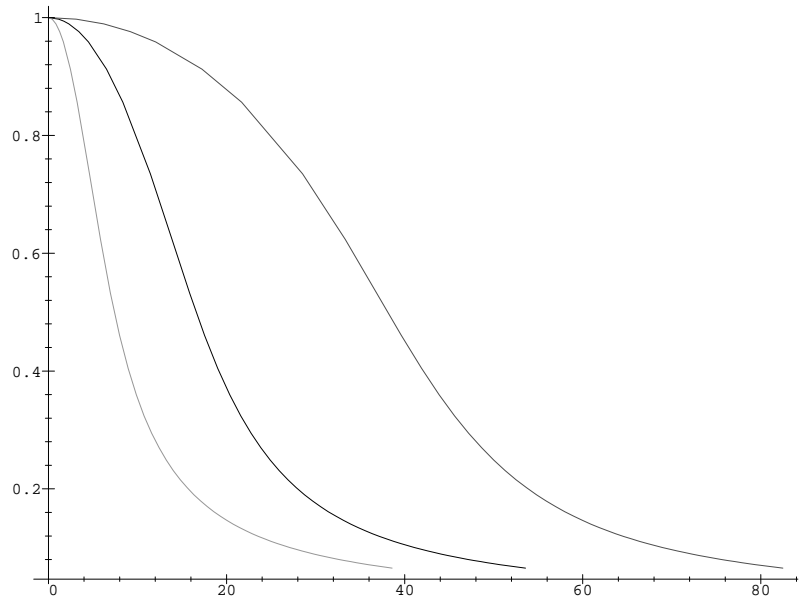}
\caption{Hubble parameter (in units of $H_0$) at times $1/H_0, 2/H_0,
  3/H_0$} 
\label{fig2}
\end{figure}

\begin{figure}[p]
\vspace{3in}
\includegraphics{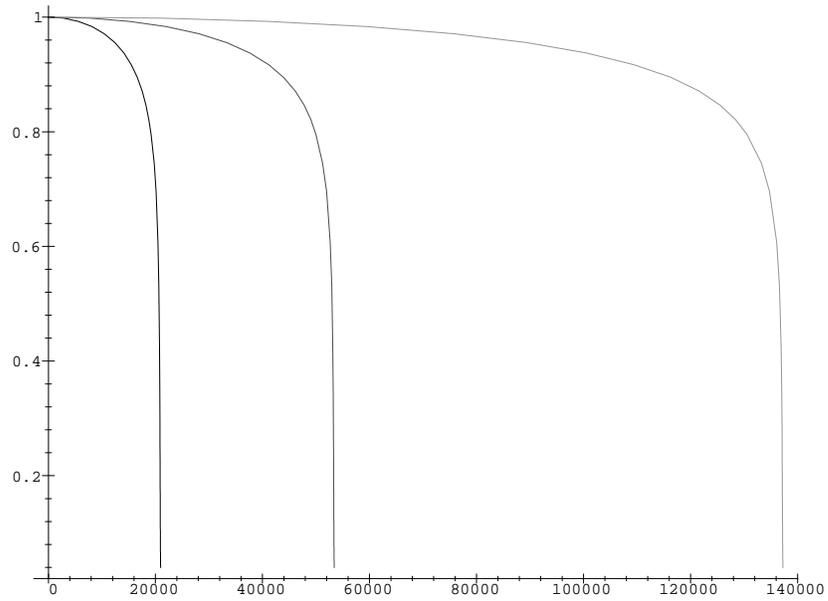}
\caption{Hubble parameter (in units of $H_0$) at times $10/H_0, 11/H_0,
  12/H_0$} 
\label{fig3}
\end{figure}

Knowing the evolution of the scale factor allows one to calculate
the temperature and energy density from pre-inflationary
quantities. This will allow us to determine these quantities at the
end of inflation and then use them as initial conditions for the
standard cosmology; hopefully giving some observable variations in
quantities to test our model.

We shall assume that before inflation there was a uniform radiation
energy density $\rh_0$ at temperature $T_c$, the critical temperature of
phase transition. We shall also assume that once the initial
transient phase of inflation is 
complete the radiation density does
not appreciably effect the evolution of the scale factor in
eq.~(\ref{scale factor}); this is motivated through observing that
Universe expansion 
dilutes $\rh$ whilst not affecting the vacuum energy density. For
radiation, $\rh$ scales as $a^{-4}$, whilst the temperature $T$ scales
as $1/a$. This yields the following two evolution equations:
\beginsubeqn
\bea
\rh(r,t) = \rh_0 \left(\frac{a_0}{a(r,t)}\right)^4 &=& \rh_o 
\exp \left(-4 t H_0 (1 - f_{\rm mon}(r)^2) \right),\\
T(r,t) = T_c  \frac{a_0}{a(r,t)} &=& T_c \exp 
\left(- t H_0 (1 - f_{\rm mon}(r)^2) \right).
\eea
\endsubeqn
As always, the spatial variations must be interpreted with the physical
distance $r'$.

We shall now estimate the required number of e-folds of inflation. To
do this we simply estimate the amount of inflation needed to expand
the monopole's core to a region of our present horizon size. This
region is in causal contact with us now, and hence the number of
e-folds of inflation must be sufficient (or more) to create a region
of this size. If there was less inflation we should have observed some
effects from the edge of the inflated core --- which are quite severe
(see later). We have not seen such effects.

The initial size of the monopole is, from eq. (\ref{ps limit}), 
$r_{\rm core} = 2 (g \et)^{-1}$. The coupling constant $g$ is known
to be about $1/50$ and the scale of symmetry breaking $\et$ is
required to be about $10^{18} \rm{GeV}$. Hence,
\beq
r_{\rm core} \sim \frac{10}{3} \x 10^{-17} {\rm GeV}^{-1}
\sim 6 \x 10^{-29} {\rm m},
\eeq
To estimate the final size of the core after inflation, we
estimate $H(r)$ to be $H(0)=H_0$; this is expected to be correct to
within a factor of two or so, and since we shall be dealing in orders
of magnitude it should prove to be a good approximation. This yields
the final core size to be
\beq
\label{d(t)}
d(t_I) = r_{\rm core} \exp \left(\sqrt{\frac{8 \pi \la \et^4}{3 m_{\rm
      pl}^2}} t_I\right) = r_{\rm core} \exp(N), 
\eeq
with $N$ the number of e-folds and $t_I$ the duration of inflationary
expansion. The size of the core $d(t_I)$ needs to be greater than the
size of our presently observable Universe, which can be estimated from
the age of our Universe, which is found from the present Hubble
constant. A standard and simple calculation gives:
\beq
d_{\rm universe} \sim \frac{6}{h} \x 10^{25} {\rm m},
\eeq
where $h \sim 0.6 - 0.8$. We define $N_{\rm crit}$ to be the minimum
number of e-folds required to produce a Universe as large as that
presently observed; the actual number of e-folds being expected
to be larger than this (though, see sec.~(\ref{sec 4.9})). Equating
$d_{\rm universe} = r_{\rm core} \exp(N_{\rm crit})$, one obtains
\beq
N_{\rm crit} = {\rm ln} (\frac{2}{h} \x 10^4), 
\eeq
and substituting above for $h$:
\beq
\label{N_crit}
N_{\rm crit} =  125,
\eeq 
which is pretty much insensitive to the value of $h$. This solves the 
horizon problem in the usual way, since the core of the
monopole is in causal contact before inflation. In addition, the
number of e-folds estimated above easily satisfies the bounds on
flatness; which requires the number of e-folds to be larger than about
sixty.

Estimating the actual number of e-folds of inflation to be about the
critical value, $N \sim N_{\rm crit}$, gives us an estimate of how the
Hubble parameter, temperature and pre-inflationary radiation density vary
with physical distance $r'$; this variation being at the point when
inflation is ending, the monopole starting to decay, but reheating not
yet occurring. Extrapolating figs. (2) and (3) to $125$ e-folds of
inflation, one sees that the variation in the Hubble parameter
with distance is
reasonably uniform upto the horizon, where a sharp falloff
(essentially to zero) takes place. Calculation verifies, with there only
being a one per cent 
variation in the Hubble parameter upto distances of $(4/5)d_{\rm
  universe}$. 

The distribution in temperature and pre-inflationary
radiation density is even more sharply defined. After $125$ e-folds
the distribution is approximately:
\beginsubeqn
\label{rho and T}
\bea
\rho(r,t_I) &=& \left\{ 
\begin{array}{ll}  0.5 \x 10^{-104} \x \rho_0 \ \ \ \  r'< d_{\rm
    universe}\\ 
                   \rho_0 \ \ \ \ \ r' \geq d_{\rm universe}
\end{array}  \right. \\ 
T(r,t_I) &=& \left\{
\begin{array}{ll}  10^{-26} \x T_c\ \ \ \  r'< d_{\rm universe}\\
                   T_c \ \ \ \ \ r' \geq d_{\rm universe}
\end{array} \right.
\eea
\endsubeqn
This massive dilution of the initial temperature and density of the
Universe renders the initial radiation density
completely unobservable. These miniscule remnants will be absorbed
into the radiation produced from reheating, of which the remnants'
contribution will be insignificant. However, conditions before
inflation are still present outside the inflated region. As we can see
from the plots the divide is immensely sharp; due to 
the scale parameter decreasing sharply to unity at the boundary of the
monopole's core.

\subsection{The End of Inflation : Decay of the Monopole.}

It is difficult to model the decay of the monopole. We have derived
eq. (\ref{phi dot}) showing how the scalar field would decay if all
the gauge fields were diluted away by inflation. However,
eq. (\ref{phi dot}) ignores 
the gravitational back reaction on the scalar field. Hence,
this equation is of limited use --- only in showing the connection
between topology and eternal inflation.

The arguments given in sec.~(\ref{sec 4.2}) indicate how a long lived 
embedded monopole may occur. Qualitatively, when the monopole decays
there is one of two ways it can do it:
\begin{itemize}
\item
Quickly. The embedded monopole lives in a long-lived stationary state
and then all points decay quickly simultaneously. This could happen
naturally by rolling, or could also be due to quantum tunnelling
processes.
\item
Slowly. Points on the edge of the core decay before points at the
centre, with an appreciable time lag. Should create appreciable large
scale inhomogeneities in cosmological parameters.
\end{itemize}
We shall assume the quick rolling scenario for simplicity. However, it
should be noted that large scale homogeneities in the slow decay
scenario would be severely decreased by later Universe evolution.

\subsection{Reheating}

In all models of inflation there is reheating, where the vacuum energy
(that does not scale with expansion) is converted into a radiation
energy density which is the origin of (virtually) all matter we see
today. The duration of reheat determines the temperature ({\em i.e.}
efficiency) produced. Owing to our model being inhomogeneous, one may
expect variations in the local energy density and temperature.

The mechanism of reheating is unchanged for our model; the only
difference being that not all points in the Universe are necessarily
required  to reheat at the same time, or have the same vacuum
energy. As usual, we shall neglect spatial gradients, so as to
approximate by local standard cosmologies.

We shall use the old mechanism of reheat; not the new more
sophisticated one (where one treats properly parametric resonance of
the scalar field) \cite{Lind94a}. We are motivated into this choice
by the comparitive simplicity, and, of course, a proper theory of
reheat should consider the theory of parametric resonance.
The scalar field decays into coherent
oscillations around the bottom of the potential well, then these
oscillations decay into radiation. The rate of oscillation decay
determines the rate of reheating.

Assuming the dominant decay width of the Higgs is into the most
massive particle --- namely, the Lepto-quark bosons --- the
decay width may be estimated to be
\beq
\Ga_\ph \sim g^2 m_{\rm LQ} = g^3 \et.
\eeq
Then the decay time of the coherent oscillations, which is the reheat
time, is about $\Ga^{-1}$. To properly estimate this one should
compare it to the Hubble expansion time at the end of inflation
$H^{-1}$. The ratio of these two quantities gives the number of Hubble
expansion times for reheating; which is
\beq
\frac{\Ga_\ph^{-1}}{H(r,t_I)^{-1}} \sim \sqrt{\frac{5 \pi}{3}} \left(
\frac{\et}{m_{\rm pl}} \right) \frac{\sqrt{\la}}{g^3} (1- f_{\rm
  mon}(r)^2).
\eeq
If $\Ga_\ph^{-1}/H(r,t_I)^{-1} \ll 1$, reheat is practically
instantaneous to a temperature 
\beq
T_{\rm RH} \sim \left( \frac{45}{4 \pi^3 g_*} \right)^{1/4} (H(r,t_I)
m_{\rm pl})^{1/2}.
\eeq
For $\Ga_\ph^{-1}/H(r,t_I)^{-1} \gg 1$, reheat is long compared to the
relevant Hubble expansion rate; the Universe then expands as if matter
dominated for the time $\Ga^{-1}$, after which the oscillations are
quickly damped --- producing a reheat temperature
\beq
T_{\rm RH} \sim \left( \frac{30}{\pi^2 g_*} \right)^{1/4} (\Ga_\ph
m_{\rm pl})^{1/2}.
\eeq
One should note that this is independent of the Hubble parameter.

Estimating parameters to be $g \sim 1/50$, $\et/m_{\rm pl} \sim 1/10$,
one finds the rate of reheat to be dependent on $\la$ only with the two
cases:
\beginsubeqn
\bea
\la < O(10^{\rm -10}) &\Rightarrow& \ {\rm FAST\  REHEAT}
\ {\rm with}\  T_{\rm RH} \sim \frac{\eta}{2} \la^{1/4} (1 - f_{\rm
  mon}(r)^2)^{1/2} \\
\la > O(10^{\rm -10}) &\Rightarrow& \ {\rm SLOW\  REHEAT}
\ {\rm with}\  T_{\rm RH} \sim 4 \x 10^{15} {\rm GeV}.
\eea
\endsubeqn

There are two notes that one should make. Firstly, for a fast reheat
there is variation with $r$ 
(distance from the centre of the core), whereas for slow reheat there
is not. Secondly, in this model it is difficult to obtain a low
temperature of reheat.

\subsection{Evolution of an Inhomogeneous Universe}

It is necessary, before the discussion on large scale inhomogeneities
after inflation, to discuss how the Universe evolves with a
non-uniform energy density. We shall model the
evolution of such a Universe by the usual approximation: neglecting
spatial gradients and treating as locally FRW.

There are two cases necessary to discuss: radiation dominated or
matter dominated. The two cases differ by an equation of state, where
the pressure is related to the energy density,
\beginsubeqn
\bea
P &=& \frac{1}{3} \rho,\ \ \ \ \ \ \ (RADIATION) \\
P &=& 0.\ \ \ \ \ \ \ \ \ \ (MATTER)
\eea
\endsubeqn
The relevance of the different states being that the
evolution of coherent oscillations of the Higgs field during
reheating is matter dominated. The evolution of the decay products
being radiation dominated.

Supposing the time at the end a previous era of Universe evolution is
$t_0(r)$, with an 
energy density (matter or radiation) of 
\beq
\rho(r', t_0) = \rho_{\rm RH}(r', t_0) + \rho_{\rm in}(r', t_0),
\eeq
the two terms representing the contributions from energy density
from reheating (respectively either a radiation density or a matter
density in the form of coherent oscillations), and the initial
radiation density of the Universe. The initial radiation density was
described in section (\ref{sec 4.3}) --- being approximately zero
inside the core, and GUT scale outside. 

In modelling the evolution of the Universe we shall ignore the
effects of $\rho_{\rm in}$ for the following reason. The initial
radiation density has the appearance of a wall; on our side it is
approximately zero and on the other side it is of GUT scale. This wall
would move towards us at the speed of light. Effects on the scale
factor and density are inside the Cauchy surface of such a wall,
which coincide with the edge of the wall itself. In essence, since the
wall moves at (approximately) the speed of light, no effects can be
seen until it hits you. This initial radiation density becomes, however,
important at very late times --- as we shall discuss in the
sec. (\ref{sec 4.9}).

\subsubsection{Radiation Dominated}

Neglecting spatial gradients, the evolution equations for the
scale factor are locally FRW, with $k=0$,
\beginsubeqn
\bea
\left( \frac{\dot{a}}{a} \right)^2 &=& \frac{8 \pi G \rho_{\rm RH}(r,
t_0)}{3} \left( \frac{a(r',t_0)}{a(r,t)} \right)^4 \\
2 \frac{\ddot{a}}{a} + \left( \frac{\dot{a}}{a} \right)^2 &=& 
\frac{8 \pi G \rho_{\rm RH}(r,t_0)}{3} \left( \frac{a(r',t_0)}{a(r,t)}
\right)^4 . 
\eea
\endsubeqn
Normalising $a(r',t_0)=1$, we obtain the following solution
\beq
a(r',t) = \left( 2 H(r,t_{\rm RH}) ( t - t_0) + 1 \right)^{1/2},
\eeq
with the consistency relation that the Universe is locally flat ({\em
  i.e.} we may neglect spatial curvature) when
that area of Universe changes its equation of state:
\beq
\label{local flat}
H(r', t_0)^2 = \frac{8 \pi G}{3} \rho(r, t_0),
\eeq
which is essentially conservation of energy.

Thus we may calculate the evolution of the Hubble constant and energy
density to be, inside the core of the monopole:
\beginsubeqn
\label{radiation dominated}
\bea
H(r',t) = \frac{\dot{a}}{a} &=& \frac{H(r,t_0)}{1 + 2
  H(r,t_0)(t-t_0)}, \\
\rho(r,t) &=& \frac{\rho(r,t_0)}{\left( 1 + 2
  H(r,t_0)(t-t_0) \right)^2}.
\eea
\endsubeqn

It is important to note that at late times the initial distribution of
the Hubble parameter is homogenised. If one considers $(t-t_0) \ll 1/H$
the usual expression for the Hubble parameter is obtained, {\em
  i.e.} $H(r',t) \sim 1/(2(t-t_0))$.

\subsubsection{Matter Dominated}

The analysis follows through exactly as for the last section: we
approximate by locally FRW Universes, neglecting spatial gradients. We
also assume the Universe is still spatially flat ($k=0$), which is to
be expected after $120$ e-folds of inflation.

Initially the Hubble constant is $H(r, t_0)$ and the matter density is
$\rho(r, t_0)$. The Universe is supposed to be dust filled, with
zero pressure. Thus, the local FRW equations are:
\beginsubeqn
\bea
\left( \frac{\dot{a}}{a} \right)^2 &=& \frac{8 \pi G \rho_{\rm RH}(r,
t_0)}{3} \left( \frac{a(r',t_0)}{a(r,t)} \right)^3 \\
2 \frac{\ddot{a}}{a} + \left( \frac{\dot{a}}{a} \right)^2 &=& 0.
\eea
\endsubeqn
Normalising $a(r',t_0)=1$, we obtain the following solution
\beq
a(r',t) = \left(\frac{3}{2} H(r,t_0) ( t - t_0) + 1 \right)^{2/3},
\eeq
with the consistency relation that the Universe is locally flat when
that area of Universe changes its equation of state:
\beq
H(r', t_0)^2 = \frac{8 \pi G}{3} \rho(r, t_0),
\eeq
Then the evolution of the matter density and the Hubble constant may
be calculated to be:
\beginsubeqn
\label{matter dominated}
\bea
H(r',t) = \frac{\dot{a}}{a} &=& \frac{H(r,t_0)}{1 + (3/2)
  H(r,t_0)(t-t_0)}, \\
\rho(r,t) &=& \frac{\rho(r,t_0)}{(1 + (3/2)
  H(r,t_0)(t-t_0))^2}.
\eea
\endsubeqn

As above, at late times the initial distribution of
the Hubble parameter is homogenised. If one considers $(t-t_0) \ll 1/H$
the usual expression for the Hubble parameter is obtained, {\em
  i.e.} $H(r',t) \sim 2/(3(t-t_0))$.

\subsection{From de-Sitter to Radiation Dominated Evolution}

There are essentially
two processes, which may be quick or slow, that determine large scale
inhomogeneities in the distributions of the Hubble constant,
energy density and temperature after inflation.
Firstly, the decay of the monopole: which may be quick or slow. 
Secondly, the duration of reheat (which depends upon the parameter
$\la$); in slow reheat the universe evolves
with matter dominating coherent oscillations of the Higgs field before
the radiation dominated era.

Hence there are a total of four different situations. For simplicity,
we shall assume the monopole decays quickly (i.e all points
simultaneously) and deal with the two cases of quick and slow reheat.

If the monopole were to decay slowly, any non-simultaneity of decay
would be transmitted into inhomogeneities of the cosmological
parameters. We do not discuss such a case.

\subsubsection{Quick Monopole Decay; Fast Reheat}

This is the simplest case and is valid for $\la < O(10^{-10})$. It is  
possible to implement this in both SUSY and NON-SUSY field
theories. The cosmology goes like:

\begin{itemize}
\item Quick monopole decay, freezing in variations of the
  Hubble parameter, energy density and temperature.
\item Universe reheats quickly from a 
  rapid decay of coherent Higgs oscillations into radiation,
  reheating the Universe.
\end{itemize}

We now expand upon this sketch, putting values to the variables.
To obtain a Universe at least as large as observed today one needs at
least $125$ e-folds of inflation (see eqs. (\ref{d(t)}) -
  (\ref{N_crit})). For simplicity, assume that this is 
the exact amount of e-folding performed. Also assume that the quartic
Higgs coupling $\la$ is in the range $10^{-9} - 1$, taking these outer
two values for computing. In addition we shall take $\et/m_{\rm pl}
\sim 1/10$.

Then at the end of inflation the Hubble constant is (from
eq. (\ref{hubble parameter})): 
\beq 
\frac{H(r,t_{\rm inf})}{1 - f_{\rm mon}(r)^2} = 
\frac{3 \la^{1/4} \et}{10} {\rm GeV} = 5 \x 10^{39} - 5 \x 10^{41}
{\rm ms^{-1}m^{-1}}.
\eeq
The distribution is very sharp
because the physical distance scale scales sharply, being related to
$r$ by $r'(r,t_{\rm
  inf})= \int_0^r \exp \left( 125 (1 - f_{\rm mon}(x)^2 )\right)
dx$.  

The scaling of the 
pre inflationary radiation density is even more severe: having a
miniscule temperature of about $10^{-26} \x T_c {\rm GeV}$
inside the inflated region and about $T_c$ outside.

The monopole decays quickly, freezing in the above variation of the
Hubble parameter. At all points in space the Higgs field falls down
into its potential well, oscillating around the bottom. These
oscillations are coherent and decay quickly (for the parameters
considered here) to a non-uniform temperature:
\beq 
T_{\rm RH} \sim \left\{ \begin{array}{ll} 
\frac{\eta}{2} \la^{1/4} (1 - f_{\rm mon}(r)^2)^{1/2}
\ \ \ \  r'< d_{\rm universe}\\ 
T_c \ \ \ \ \ r' \geq d_{\rm universe}
\end{array}  \right.
\eeq
and by eq.~(\ref{local flat}) we determine the density to be:
\beq
\rho(r,t_{\rm inf}) = \left\{ \begin{array}{ll} 
\frac{3 H_0^{1/2}}{8 \pi G} (1 - f_{\rm mon}(r)^2)^{1/2},
\ \ \ \  r'< d_{\rm universe}\\ 
\rho_c \ \ \ \ \ r' \geq d_{\rm universe}
\end{array}  \right.
\eeq 
which scales the same as temperature.

The Universe then evolves as radiation dominated, according to
eq.~(\ref{radiation dominated}).

\subsubsection{Quick Monopole Decay; Slow Reheat}

This case is valid for $\la < O(10^{-10})$.

\begin{itemize}
\item Quick monopole decay, freezing in variations of the
  Hubble parameter, energy density and temperature.
\item A long period of reheat, where the Universe evolves
  with coherent Higgs oscillations defining the equation of state,
  which is matter dominated. The matter dominated evolution
  homogenises the large scale variations in cosmological
  parameters. 
\item A rapid decay of coherent Higgs oscillations into radiation,
  reheating the Universe. 
\end{itemize}
We complement this sketch with a more detailed examination of the
transition.

The treatment follows as above for the decay of the monopole and hence
for the initial distribution of the Hubble parameter. However, now the
coherent Higgs oscillation take a long time to decay and the Universe
evolves as matter dominated for a time $t \sim \Gamma_\phi^{-1} = (g^3
\et)^{-1}$. Using eq. (\ref{matter dominated}), one determines the
Hubble parameter to vary 
as:
\bea
H(r, t_{\rm RH}) &=& \frac{H_0 (1 - f_{\rm mon}(r)^2)}{1 + \frac{3
    H_0}{g^3 \et} (1 - f_{\rm mon}(r)^2)}, \\
{\rm with}\ \ \ \ \frac{3 H_0}{g^3 \et} &\sim& 5 \x 10^5 \sqrt{\la}.
\eea
Hence as long as $\la > O(10^{-10})$, the effects of the Higgs
oscillations are important on the variation of the Hubble
parameter. For $\la \gg O(10^{-10})$ these effects homogenise the
Hubble parameter to the value
\beq
H(r, t_{\rm RH}) \sim \frac{2}{3} g^3 \et \sim 5 \x 10^{12} {\rm
  GeV}.  
\eeq

Then the decay of coherent Higgs oscillations is quick, leading to a
reheat temperature
\beq
T_{\rm RH} \sim \frac{2}{5} (g^3 \et m_{\rm pl})^{1/2}
           \sim 5 \x 10^{15} {\rm GeV}.
\eeq

\subsection{Constraints from Gravitational Radiation}

It is well known that Planck scale inflationary models produces
copious amounts of gravitational radiation. This gravitational
radiation gives a contribution to the cosmic microwave
background \cite{Ruba82}. Thus COBE anisotropy measurements constrain 
such inflationary models.

The model of inflation presented in this paper is locally similar
to standard models of vacuum dominated inflationary models; though
with the vacuum characterised by the energy density
\beq
\epsilon = \la \et^4 (1 - f_{\rm mon}(r)^2)^2.
\eeq
Hence, from \cite{Ruba82} the contribution to the quadrupole moment of the
cosmic microwave background is
\beq
\left( \frac{\De T}{T} \right)_{\rm quad} = 2.4 \frac{\la
    \et^4}{m_{\rm pl}^4} (1 - f_{\rm mon}(r)^2)^2.
\eeq
COBE gives the observed quadrupole anisotropy to be $(\De T / T)_{\rm
  quad} \sim 7 \x 10^{-6}$. Estimating $(\et/m_{\rm pl}) \sim 0.1$
yields the following constraint on $\la$
\beq
\la < 3 \x 10^{-2}.
\eeq

It should be noted that there may be other contributions to the
quadrupole anisotropy that may further constrain $\la$.

\subsection{The Universe at Very Late Times}
\label{sec 4.9}

Examining eq. (\ref{rho and T}), one sees that the divide
between the 
inflated region of space and the uninflated GUT scale regions is
extremely sharp; being completely due to the scaling of the distance
scale, eq. (\ref{scaling}). One could also interpret this as a massive
time dilation as one passes from present day physics to GUT-scale
physics across this divide. 

The upshot of this is that this scenario for inflation predicts a
region of space akin to our own, but surrounded by a wall of
GUT scale energy. Across the wall there is a massive pressure
difference caused by the large density (and temperature) variation.
Such a wall would move into the inflated region at practically
the speed of light.

We have shown previously (eqs. (\ref{d(t)}) - (\ref{N_crit})) that in
order to create a 
region of space at the end of inflation large enough to contain our
presently observable Universe ({\em i.e.} the wall has not hit us) the
number of e-folds of inflation has to be greater than $N_{\rm crit}
=125$. The question is, therefore, given that at least $125$ e-folds
of inflation has taken place, how many more e-folds can we 
expect --- {\em i.e.} how much longer do we have before this wall of
energy hits us?

Recall that the number of e-folds of inflation determines the size of
the inflated region by (we are talking about the diameter,
assuming the region to be spherical)
\beq
d_{\rm universe} = d_{\rm core} e^N,\ {\rm where}\ N = H_0 t_{\rm
  decay},
\eeq
where the core radius $d_{\rm core} = 2 (g \et)^{-1}$, $H_0$ is the
Hubble parameter of eq. (\ref{scale factor}) and the decay time of the
monopole is 
probabilistically distributed as (from eq. (\ref{probability}))
\beq
P(t_{\rm decay}>t) = e^{-\frac{t}{\al}}.
\eeq
The expected (mean) time of
decay is $\bar{t} = \al$. For small times $\al \sim \et^{-1}$
(determining whether inflation initiates). After inflation initiates,
the characteristic time scale is the Hubble time 
$H_0^{-1}$. Thus for large times we estimate $\al \sim H_0^{-1}$ and
the above distribution is approximately Poisson, as long as we are
conditional on the monopole not decaying quickly.

Given the number of e-folds of inflation is greater than $125$,
the Poisson nature of the probability distribution at large times gives
the conditional probability for the number of e-folds:
\beq
P({\rm e-folds}>N \ |\  N>N_{\rm crit}) = 
\exp \left( \frac{N-N_{\rm crit}}{H_0 \al} \right).
\eeq
This yields the expected number of e-folds of inflation conditional on
$N> N_{\rm crit}$ to be
\beq
\bar{N} = N_{\rm crit} + 1.
\eeq

Hence if one estimates the age of our Universe (the time since
inflation finished) to be about $10-15$Gyr, from which the size of our
observable universe is about $10^{26}$m. Then the actual size of
inflationary universe created in this model of inflation is
probabilistically distributed with a mean of $e$ times this. 
Before estimating the amount of time left, one should
remember that we are not necessarily at the `center' of the Universe,
but somewhere randomly distributed in the Universe that has not seen
the wall of energy coming towards us --- this gives a factor
of half
from our expected time. Hence we can realistically expect about
$(e-1)/2 \x 10-15$Gyrs, {\em i.e.} about $8-13$Gyrs left
\footnote{
It is amusing to note that if one used $\et^{-1}$ instead as
$H_0^{-1}$ as the characteristic time scale then one obtains a value
$10^{-4}$s as the expected time}.


\section{Conclusions}
\label{sec 5}

By nature, the work in this paper is open ended. Hence, in this
section we shall briefly discuss some directions in which the work
could be extended. We shall briefly discuss the 
consequences of non-spherically symmetric decay;
realisation with other unstable defects; and the cause and
observation of large scale inhomogeneities from such scenarios.

However, firstly we shall give a brief summary of the features of
non-topological inflation as presented in this paper.

\subsection{Non-topological Inflation}

We consider a Yang-Mills field theory that realistically describes a
GUT. Our theory must have monopole solutions
\bea
\Ph(r,\th) &=& f_{\rm mon}(r) \underline{\hat{r}},\nn \\
A(r,\th) &=& \frac{g_{\rm mon}(r)}{qr} \ep_{\mu a b} T_b,\nn
\eea
where $q$ is the order of the coupling constants and $f_{\rm mon}(r)$
and $g_{\rm mon}(r)$ are the profile functions of the
monopole. Monopole solutions may be either stable or unstable.

\begin{itemize}
\item
Coupling the monopole solution to gravity yields inflation provided
\[
\left( \frac{\eta}{m_{\rm pl}} \right)^2 > \frac{1}{4 \pi}.
\]
Additionally $\la$ is constrained so that
\[
\frac{1}{2} \la \left( \frac{\et}{m_{\rm pl}} \right)^2 < 1, 
\]
otherwise the monopole becomes a black hole.
\item 
The inflating monopole solution has a Hubble parameter distribution of 
\bea
H(r,t) = \frac{\dot{a}}{a} = H_0 \left(1 - f_{\rm mon}(r)^2 \right), 
\nn \\
{\rm where}\ \  H_0 = \sqrt{\frac{8 \pi \la \et^4}{3 m_{\rm pl}^2}}.
\nn
\eea
and the physical coordinate $r'$ is related to $r$ by the scaling
relation $r'(r,t) = \int_{x=0}^{r} \frac{a(x,t)}{a_0} {\rm d}x$, with
$a$ the scale factor which, during inflation, is of the form $a(r,t) =
a_0 \exp(H_0 t (1 - f_{\rm mon}(r)^2))$.
\item
Topologically stable monopoles inflate eternally: difficult to
reconcile with the present state of the Universe. Unstable monopoles
decay probabilistically: stipulate that we lived in such a monopole
that (randomly) lived long enough for us to exist.
\item
For the Universe to be as large as we see today 
(the diameter of the Universe $d_{\rm universe} \sim 10^{26}m$)
one requires the decay
time $t_{\rm decay} > 125 H_0^{-1}$. After such time the distribution
of initial density and temperature is:
\bea
\rho(r,t_I) &=& \left\{ 
\begin{array}{ll}  0.5 \x 10^{-104} \x \rho_0 \ \ \ \  r'< d_{\rm
    universe}\\ 
                   \rho_0 \ \ \ \ \ r' \geq d_{\rm universe}
\end{array}  \right. \nn \\ 
T(r,t_I) &=& \left\{
\begin{array}{ll}  10^{-26} \x T_c\ \ \ \  r'< d_{\rm universe}\\
                   T_c \ \ \ \ \ r' \geq d_{\rm universe}
\end{array} \right.\nn
\eea
and the form of the Hubble parameter is also very sharp. Here $r'$ is
the physical radial coordinate.
\item
The monopole then decays. Resulting reheat (from coherent oscillations of
the scalar field about the bottom of the potential) slow if $\la >
O(10^{-10})$ and fast otherwise. Temperature of reheat distributed as:
\bea
{\rm FAST\  REHEAT} \ &{\rm with}&\  T \sim
 \left\{ \begin{array}{ll} 
\frac{\eta}{2} \la^{1/4} (1 - f_{\rm mon}(r)^2)^{1/2}
\ \ \ \  r'< d_{\rm universe}\\ 
T_c \ \ \ \ \ r' \geq d_{\rm universe}
\end{array}  \right. \nn \\ 
{\rm SLOW\  REHEAT}
\ &{\rm with}&\  T \sim
 \left\{ \begin{array}{ll} 
4 \x 10^{15} {\rm GeV} \ \ \ \  r'< d_{\rm universe}\\ 
T_c \ \ \ \ \ r' \geq d_{\rm universe}
\end{array}  \right. \nn 
\eea
where $r$ is related to the physical coordinate $r'$ by the scaling
relation above.
\item
After reheating the Hubble parameter is distributed as:
\bea
{\rm FAST\  REHEAT} \ &{\rm with}&\  H(r,t_{\rm RH}) \sim
 \left\{ \begin{array}{ll} 
\frac{3 \la^{1/4} \eta}{10} (1 - f_{\rm mon}(r)^2)
\ \  r'< d_{\rm universe}\\ 
0 \ \ \ \ \ r' \geq d_{\rm universe}
\end{array}  \right. \nn \\ 
{\rm SLOW\  REHEAT}
\ &{\rm with}&\ H(r,t_{\rm RH}) \sim
 \left\{ \begin{array}{ll} 
\frac{2}{3} q^3 \eta \ \ \ \  r'< d_{\rm universe}\\ 
0 \ \ \ \ \ r' \geq d_{\rm universe}
\end{array}  \right. \nn 
\eea
where, as usual, one uses the scaled distance.
\item
For inflationary gravitational radiation to be less
than present bounds, one requires:
\[
\la < 3 \x 10^{-2}. 
\]
\item
The divide between inflationary and non-inflationary region is very
sharp and 
moves into the inflationary region at (practically) the speed of
light. Given the Universe is at least as large as we see, the
conditional expectation value for the amount of time left before the
wall reaches us is around $8 - 14$Gyrs.
\end{itemize}

\subsection{Non-Spherically Symmetric Decay}

In the above treatment of inflation, we have always assumed that our
monopole decays through a spherically symmetric mode. Hence, the
distribution of cosmological parameters has also spherical symmetry
around the centre of the monopole.

However, there are other decay modes as well as the
spherically symmetric one. Higher decay modes
correspond to higher spherical harmonics, and although they correspond 
to higher energies, the route by which the monopole decays
depends upon the initial conditions of the monopole --- which are
random. The spherical mode we have been using corresponds to the
lowest energy mode $Y_{00}(\th, \phi)$, higher energy decay modes
correspond to $Y_{lm}(\th, \phi)$ harmonics. The lowest harmonic that
is non-spherically symmetric is the $Y_{1m}(\th, \phi)$, with
$m=-1,0,1$, and gives a dipolar decay --- resulting in an ellipsoidal
inflated region.

In general one may expect that the decay is the linear superposition
of several decay modes.

\subsection{Realisation with other Defects}

Although the treatment, so far, has only been concerned with embedded
monopoles, it does seem likely that it may also be realisable with
other defects --- such as embedded vortices or embedded domain walls.
{\em The} important criterion for inflation is longevity of the
defect, which we argue can be due to the defect having a corresponding
saddle point in configuration space. Embedded vortices and embedded
domain walls also have saddle points in configuration space, and are
thus also inflationary candidates.

Each type of defect gives a different symmetry of cosmological
parameters: an embedded vortex should give rise to cylindrical
symmetry (with superimposed harmonic decay modes) and an embedded
domain wall would give rise to cosmological parameters with
reflection symmetry.

\subsection{Possible Observable Consequences}

From our results on the distribution of cosmological parameters after
inflation, one can see that the parameters (temperature, density,
Hubble parameter) decrease towards the edge of the inflated
region. Since the decrease is so small and is so close to the edge of
the inflated region such inhomogeneities could only be manifest in
quantities that originated a very long time ago, just reaching us
now. Possible candidates are either a decrease in the Cosmic Microwave
Background temperature (unlikely to be detectable because of
uncertainty in the Hubble constant now); or, more likely, the Cosmic
Neutrino Background temperature (which decouples at about $10^{-9}$
seconds). Since the Cosmic Neutrino background has never been
observed, this is, of course, only a speculation.

It should be noted that a non-spherically symmetric Universe (either
from higher monopole decay modes or by using other defects) would give
a non-spherically symmetric ({\em i.e.} non-uniform) contribution to
the Cosmic microwave and Neutrino backgrounds.

\bigskip
\bigskip


{\noindent{\Large{\bf Acknowledgements.}}}

\nopagebreak

\bigskip

\nopagebreak

This work is supported in part by PPARC. We acknowledge
EPSRC for research studentships. We wish to thank A.C. Davis
and M. Trodden for interesting discussions related to this work.

\bigskip
\bigskip


{\noindent{\Large{\bf Appendix: The Non-existence of Sphalerons in
      Flipped-$SU(5)$ }}}

\nopagebreak

\bigskip

\nopagebreak

For the later parts of this paper it is important to know what
monopole-type configurations are solutions to flipped-$SU(5)$. In
\cite{me1} the motivation for studying V-strings in flipped-$SU(5)$
came from an analogy in structure between flipped-$SU(5)$ and the
Weinberg-Salam model. Carrying the analogy further leads to a
consideration of the existence of Sphaleron configurations. We shall
now show that, rather surprisingly, such a configuration is not a
solution to flipped-$SU(5)$. 

If a Sphaleron configuration was a solution to flipped-$SU(5)$ then an
embedded subtheory could be constructed which contains it. The form of
the embedded subtheory has to be the same as the Weinberg-Salam
model. Namely $G_{\emb} = SU(2) \x U(1)$, with $H_{\emb} = H \cap
G_{\emb} = U'(1)$, where $U'(1)$ is generated by a linear combination
of a generator of $U(1)$ and $SU(2)$. A little thought shows that
$G_{\emb} \subset SU(5) \x \widetilde{U(1)}$ in the following way:
\bea
SU(5) &\x& \widetilde{U(1)} \nn \\
\cup \ \ &\ &\ \| \\
SU(2) &\x& U(1) \nn 
\eea
with $H_{\emb} = U(1)_Y$. For the breaking $G_{\emb} \rightarrow
H_{\emb}$ to properly mirror that of the Weinberg-Salam model the
generators of the $SU(2)$-algebra $\{T, T', [T,T']\}$ must be such that 
\beq
T, T' \in \underline{m}_2,\ \ \ {\rm with} \ \ \ [T, T'] = T^{15}.
\eeq
It is this condition that cannot be satisfied.

As $\{T, T', T^{15}\}$ generate an $SU(2)$-algebra, the following
relation holds:
\beq
[T,[T,T^{15}]]=T^{15}.
\eeq
Taking $T$ to be defined as in eq. (\ref{m}) yields the following
relations:
\beginsubeqn
\bea
\underline{A} \underline{A}^\dagger &=& \frac{1}{5} {\bf 1}_3, \\
\underline{A}^\dagger \underline{A} &=& -\frac{3}{10} {\bf 1}_2, 
\eea
\endsubeqn
which gives
\beq
\underline{v}_i^\dagger \underline{v}_j = \frac{1}{5} \de_{ij}.
\label{eq-19}
\eeq
Since $\underline{v}_i$ is a two-component complex vector,
eq. (\ref{eq-19}) says that we must find {\em three} orthogonal
vectors in a {\em two}-dimensional vector space. This is impossible.

\bigskip
\bigskip



\end{document}